\documentclass[twocolumn]{jpsj3}
\usepackage{bm}
\usepackage{color}

\def\runauthor{J. \textsc{Otsuki} et al.}

\title{
Effect of Disorder on Fermi surface in Heavy Electron Systems
}

\author{%
Junya \textsc{Otsuki}\thanks{E-mail address: otsuki@cmpt.phys.tohoku.ac.jp}, 
Hiroaki \textsc{Kusunose}$^1$, and Yoshio \textsc{Kuramoto}
}

\inst{%
Department of Physics, Tohoku University, Sendai 980-8578\\
$^1$Department of Physics, Ehime University, Matsuyama 790-8577
}

\recdate{\today}

\abst{%
The Kondo lattice model with substitutional disorder is studied with attention to
the size of the Fermi surface and the associated Dingle temperature. 
The model serves for understanding heavy-fermion Ce compounds alloyed with La 
according to substitution Ce$_x$La$_{1-x}$.
The Fermi surface is identified from the steepest change of the momentum distribution
of conduction electrons, and is derived at low enough temperature 
by the dynamical mean-field theory (DMFT) combined with the coherent potential approximation (CPA).
The Fermi surface without magnetic field
{\it increases} in size 
with decreasing $x$ from $x=1$ (Ce end), and disappears at such $x$ that gives the same number of localized spins as that of conduction electrons.
From the opposite limit of $x=0$ (La end), the Fermi surface broadens quickly as
$x$ increases, but stays at the same position as that of the La end.
With increasing magnetic field,  a metamagnetic transition occurs,
and the Fermi surface above the critical field changes continuously across the whole range of $x$.
The Dingle temperature takes a maximum around $x=0.5$.
Implication of the results to experimental observation is discussed.
}

\kword{%
Large Fermi surface, Dingle temperature, metamagnetism, Ce$_x$La$_{1-x}$Ru$_2$Si$_2$, 
coherent potential approximation (CPA), 
continuous-time quantum Monte Carlo (CT-QMC)
}

\begin{document}
\maketitle

\section{Introduction}
%

The Fermi surface (FS) reflects strongly the formation of heavy quasiparticles.
Suppose that 4f electrons in Ce compounds behave as localized spins decoupled with conduction electrons, the volume surrounded by the FS does not include 4f electrons in a sense of Luttinger's theorem\cite{Luttinger60}.
In this case, the topology of the FS is the same with that of the corresponding La compounds, which is referred to as ``small FS".
On the other hand, in the heavy-fermion state, the quasiparticles involve 4f electrons, which therefore contribute to the FS volume to yield ``large FS".
Thus, investigating the FS is a direct way of observing the formation of the heavy-fermion state. 

CeRu$_2$Si$_2$ is a typical material that exhibits heavy-fermion behavior.
Magnetic field gives rises to a metamagnetic crossover\cite{Sakakibara95} 
and simultaneously, the topology of the FS changes from 4f-itinerant one to that similar to LaRu$_2$Si$_2$, accompanied by an enhancement of the cyclotron effective mass.\cite{Aoki93}
According to Luttinger's theorem, the total volume surrounded by the FS cannot be changed continuously in an ideal system. 
Hence, the change of the FS topology should be referred to as ``transition",
but should be distinguished from the ordinary phase transition in thermodynamics.
A similar FS transition has been observed 
by applying pressure in some compounds such as CeRhIn$_5$\cite{Shishido05} and CeIn$_3$\cite{Settai05}.

Substitution of Ce ions with La ions is an effective technique to reveal the role of 4f electrons in physical properties.
Especially for the FS, it is useful since the topologies at both limits are different.
Strictly speaking, the momentum distribution function has no discontinuity even at $T=0$ in the alloys with disorder.
Nevertheless, de Haas-van Alphen (dHvA) oscillations have been observed for wide range of the Ce concentration.\cite{Harrison04, Goodrich99, Endo06, Matsumoto08}
Hence, in this paper, we use the term ``Fermi surface" for the steep slope in the momentum distribution.

Theoretical studies have been performed 
for binary alloys of Ce and La ions, 
and the change from the Kondo singlet in the dilute limit of Ce 
to the heavy-fermion state in the high density limit has been discussed\cite{Yoshimori-Kasai, Schlottmann92, Shiina95, Wermbter96, Miranda, Mutou, Burdin-Fulde, Li-Qiu91, Yu08, Grenzebach08, Watanabe-Ogata10}.
However, the location of the FS in the disordered system has not been derived
despite
extensive studies from experimental side.
In this paper, we theoretically investigate the effect of disorder on the FS.
In considering dHvA results, we should pay attention that dHvA experiments observe electronic states under magnetic field.
Actually, under magnetic field,  
the FS topology may look as if no 4f electrons contribute, 
even though the quasiparticle is composed mainly of 4f electrons\cite{Watanabe00, Miyake-Ikeda}.
We shall demonstrate that the substitution leads to 
qualitatively different FS
depending on strength of the magnetic field.

We examine the effect of disorder based on the Kondo lattice model.
Details of the model and the computational 
method are explained in the next section.
We then show numerical results for the case of Ce end ($x=1$) 
under magnetic field in \S\ref{sec:mag}.
Results for disordered system are given in \S\ref{sec:disorder}.
We summarize in \S\ref{sec:discussion} with discussions on realistic materials.

\section{Model and Computational Method}

We consider a disordered Kondo lattice model in which localized spins 
are randomly distributed with concentration $x$.  We refer these sites to
the A-sites. 
The Hamiltonian then reads
\begin{align}
{\cal H} &= \sum_{\bm{k}\sigma} (\epsilon_{\bm{k}}-\mu_{\sigma}) c_{\bm{k}\sigma}^{\dag} c_{\bm{k}\sigma} \nonumber \\
&+ \sum_{i \in {\rm A}} \left[ 
J \sum_{\sigma\sigma'} c_{i\sigma}^{\dag} \bm{\sigma} c_{i\sigma'} \cdot (\bm{S}_i)_{\sigma' \sigma}
+ 2H S_i^z \right], 
\end{align}
where $\sigma=\pm$, and $c_{i\sigma}=N^{-1/2}\sum_{\bm{k}} c_{\bm{k}\sigma} e^{i\bm{k}\cdot \bm{R}_i}$ with $N$ being the number of sites.
The chemical potential includes the Zeeman splitting as $\mu_{\sigma}=\mu - \sigma H$, and the $g$-factor for conduction electrons is assumed to be the same as localized spins.

A randomly disordered alloy can be dealt with by the coherent potential approximation (CPA)\cite{Yonezawa-Morigaki, Elliott}.
On the other hand, strong local correlation 
can be taken into account by the dynamical mean-field theory (DMFT)\cite{Georges}, which can be combined with the CPA\cite{Yoshimori-Kasai, Schlottmann92, Shiina95, Wermbter96, Miranda, Mutou, Burdin-Fulde, Li-Qiu91, Yu08, Grenzebach08}. 
We shall give a brief review of this framework for application to the Kondo lattice model.

We assume, for the electronic state, the translational symmetry of the lattice. 
This assumption can be realized by taking an average over the random configuration of two ions.
Within a local approximation for the self-energy, the single-particle Green function for conduction electrons is given by
\begin{align}
	G_{\sigma}(\bm{k}, z) = [z-\epsilon_{\bm{k}} + \mu_{\sigma} - \Sigma_{\sigma}(z)]^{-1}.
\label{eq:G_k}
\end{align}
The self-energy $\Sigma_{\sigma}(z)$ incorporates both the coherent potential and the many-body effect due to local interactions.
We find an optimal self-energy from the solution of a single-impurity problem in an effective medium.
We define a cavity Green function ${\cal G}_{0\sigma}(z)$ by
\begin{align}
	{\cal G}_{0\sigma}(z)^{-1} = \bar{G}_{\sigma}(z)^{-1} + \Sigma_{\sigma}(z),
\label{eq:eff_medium}
\end{align}
where $\bar{G}(z)=N^{-1}\sum_{\bm{k}} G(\bm{k}, z)$. 
We omit the argument $z$ and index $\sigma$ hereafter for simplicity.
The local Green function ${\cal G}_{\alpha}$ with $\alpha={\rm A}$, B is defined for the local moment site A and the non-magnetic site B.
It is given by
\begin{align}
	{\cal G}_{\alpha} = {\cal G}_0 + {\cal G}_0 t_{\alpha} {\cal G}_0,
\label{eq:G_AB}
\end{align}
where $t_{\alpha}$ is the 
single-site
$t$-matrix, and the interaction has been assumed to be local.

In the CPA, the cavity Green function ${\cal G}_0$ is determined so that the local Green function, 
$\langle {\cal G} \rangle$, 
averaged over two kinds of sites
is identical to that of the whole lattice, $\bar{G}$:
\begin{align}
	\bar{G} = \langle {\cal G} \rangle = x {\cal G}_{\rm A} + (1-x) {\cal G}_{\rm B}.
\label{eq:CPA_condition}
\end{align}
This CPA condition can be written in a form which does not include ${\cal G}_0$.
Using eq.~(\ref{eq:eff_medium}), eq.~(\ref{eq:G_AB}) is rewritten in terms of $\bar{G}$ as
\begin{align}
	{\cal G}_{\alpha} = \bar{G} + \bar{G} \tilde{t}_{\alpha} \bar{G},
\end{align}
where 
\begin{align}
	\tilde{t}_{\alpha} = \frac{(\Sigma_{\alpha}-\Sigma)}{1- \bar{G} (\Sigma_{\alpha}-\Sigma)},
\end{align}
with $\Sigma_{\alpha}^{-1}=t_{\alpha}^{-1}+{\cal G}_0$.
The CPA condition (\ref{eq:CPA_condition}) is then written as
\begin{align}
	\langle \tilde{t} \rangle = x \tilde{t}_{\rm A} + (1-x) \tilde{t}_{\rm B} =0,
\end{align}
which leads to
\begin{align}
	\Sigma = \frac{\langle t \rangle}{1 + {\cal G}_0 \langle t \rangle}, \qquad
	\langle t \rangle = xt_{\rm A} + (1-x)t_{\rm B}.
\label{eq:self-CPA}
\end{align}
This equation gives the self-energy in the CPA.
For $x=1$, the equation is reduced to that in the pure Kondo lattice model\cite{Otsuki-KLM1}.
The effective medium ${\cal G}_0$ is determined self-consistently from eqs.~(\ref{eq:G_k}), (\ref{eq:eff_medium}) and (\ref{eq:self-CPA}).

We solve the effective impurity problem using the continuous-time quantum Monte Carlo (CT-QMC) method adapted to the Kondo model\cite{Otsuki-CTQMC}.
We use a semi-circle density of states
$\rho(\epsilon) = (2/\pi D) \sqrt{1-(\epsilon/D)^2}$
with the band cutoff $D=1$,
and use the following numerical parameters:
$n_{\rm c}=0.9$ for the number $n_{\rm c}$ of conduction electrons per site,
and $J=0.34$.
The resultant Kondo temperature is estimated as $T_{\rm K} \sim 0.1$. 
With these parameters, the antiferromagnetic ordering is suppressed by formation of the Kondo singlet\cite{Peters-Pruschke}.

\section{Effect of Magnetic Field at $x=1$}
\label{sec:mag}

\begin{figure}[tb]
	\begin{center}
	\includegraphics[width=8cm]{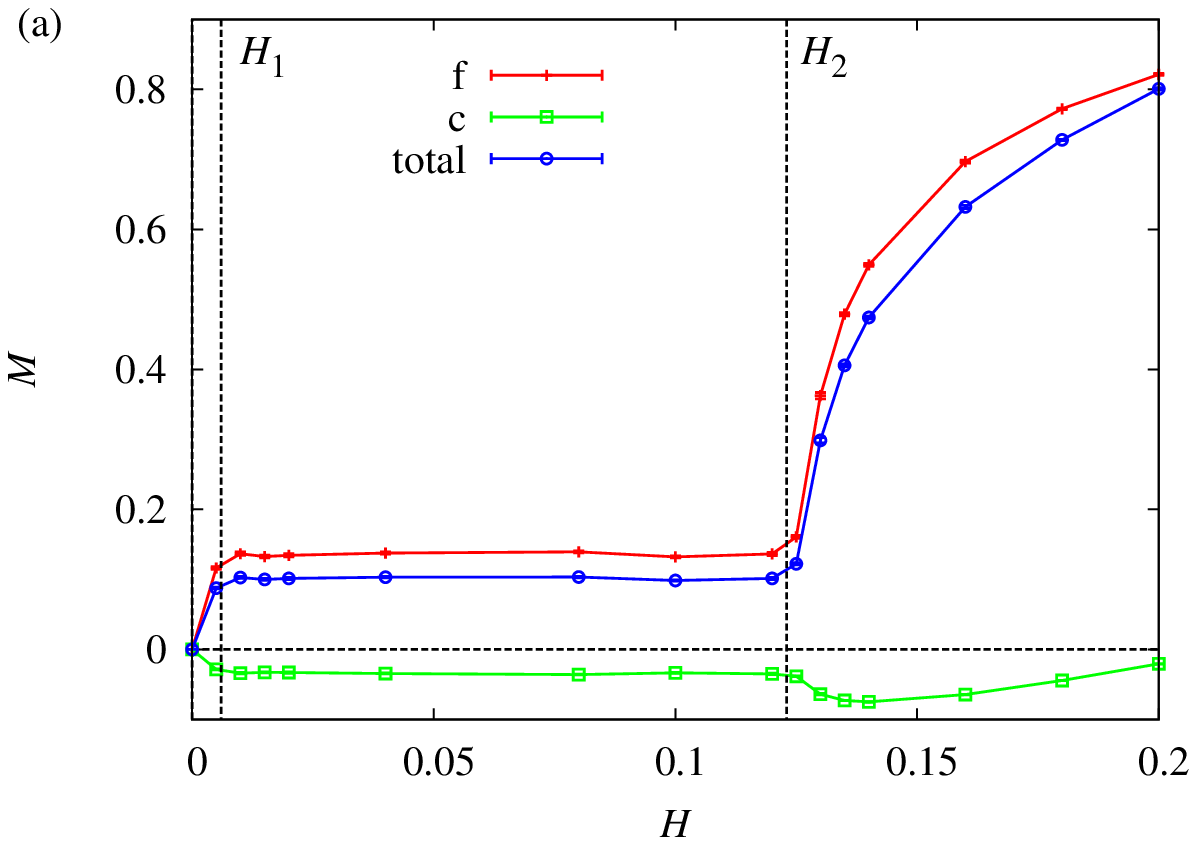}
	\includegraphics[width=8cm]{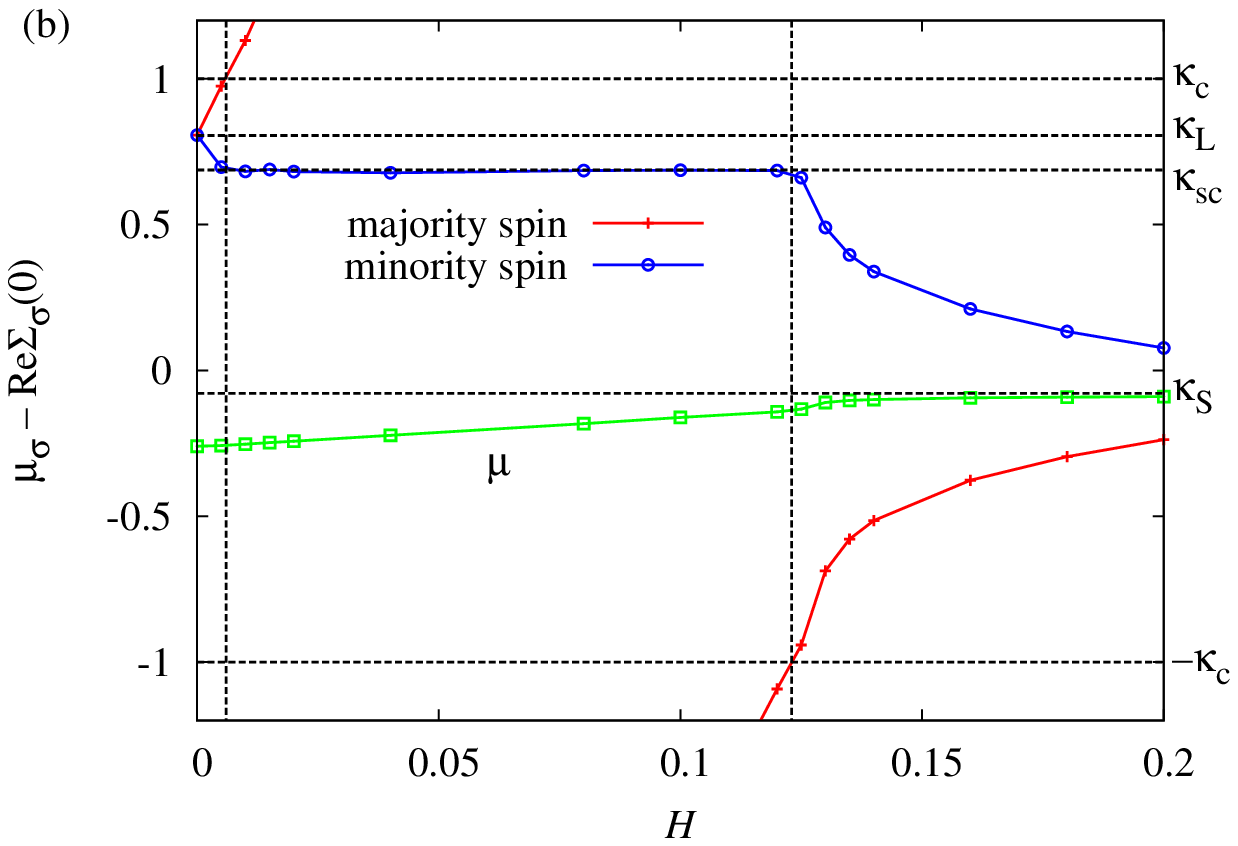}
	\includegraphics[width=8cm]{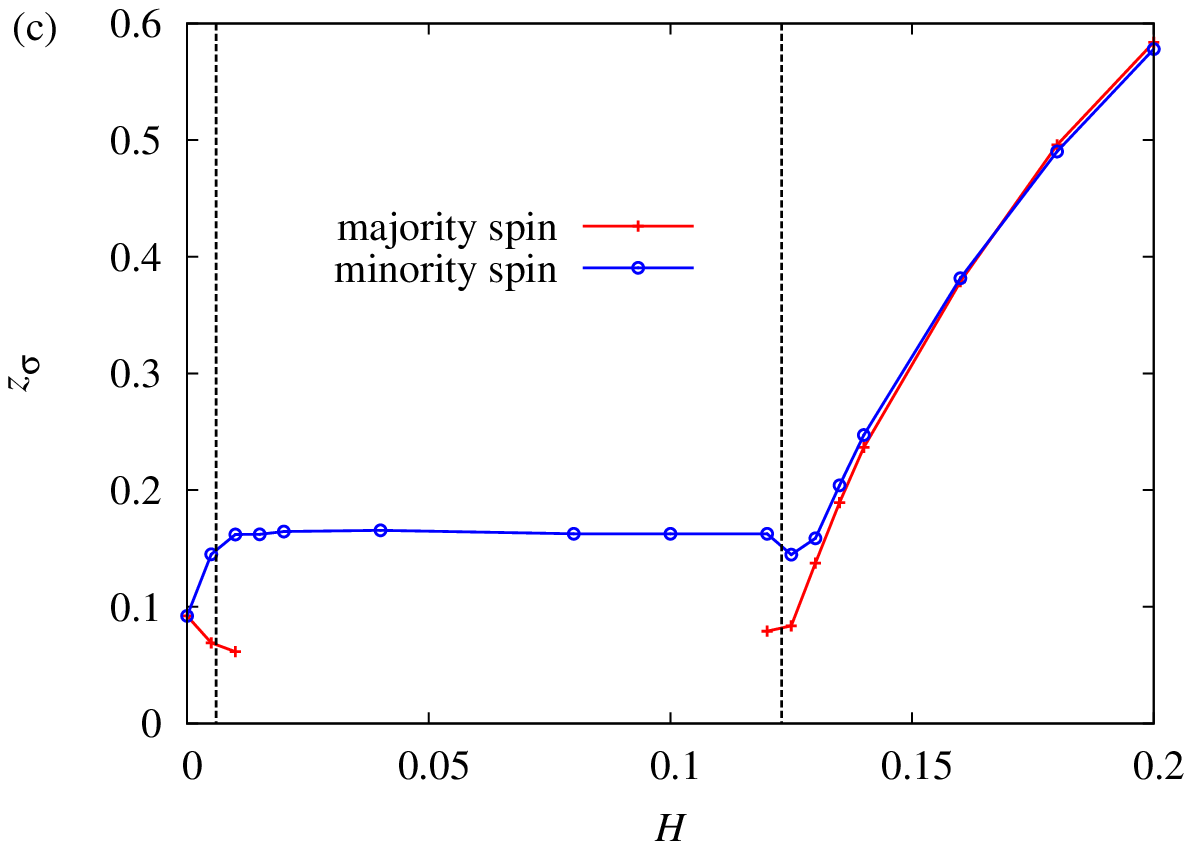}
	\end{center}
	\caption{(Color online) Magnetic field dependences of (a) magnetization $M$, (b) Fermi momentum $\kappa = \mu_{\sigma} - {\rm Re}\Sigma_{\sigma}(0)$, and (c) spin-dependent renormalization factor $z_{\sigma}$. 
The Fermi level lies in the energy gap of the majority spin band for 
$H_1<H<H_2$.
Parameters are chosen as $x=1$, $J=0.34$, $n_{\rm c}=0.9$, and $T=0.0025$.}
	\label{fig:H_depend}
\end{figure}

We begin with 
the pure system (Ce end) under magnetic field.
Figure~\ref{fig:H_depend}(a) shows the magnetization process.
We define $H_1$ and $H_2$ as lower and upper fields of the magnetization plateau, respectively.
The appearance of the magnetization plateau is explained in a quasiparticle band picture, which also gives an account of the change of the FS topology\cite{Watanabe00, Miyake-Ikeda}.
In the heavy-fermion system, an almost 
dispersionless 
quasiparticle band is located close to the Fermi level. 
Mixing with a conduction band results in a hybridization gap, which is of the order of $T_{\rm K}$ in the Kondo lattice\cite{Otsuki-LFS}.
Under the magnetic field, one of spin components is filled up to the energy gap. 
The magnetization does not change in this region, namely, $H_2-H_1 \sim T_{\rm K}$.

Let us observe the change of the FS topology.
In the DMFT, $\bm{k}$-dependence of the Green function enters only through $\epsilon_{\bm{k}}$ as shown in eq.~(\ref{eq:G_k}). 
Hence, we regard $\kappa=\epsilon_{\bm{k}}$ as ``momentum" hereafter.
The Fermi momentum can be traced by the renormalized chemical potential $\mu_{\sigma}-{\rm Re}\Sigma_{\sigma}(0)$, since the FS of each spin component appears at the momentum where $\kappa=\mu_{\sigma}-{\rm Re}\Sigma_{\sigma}(0)$ is satisfied\cite{Otsuki-LFS}.
Figure~\ref{fig:H_depend}(b) shows $\mu_{\sigma}-{\rm Re}\Sigma_{\sigma}(0)$ as a function of $H$.
There are three regions, and correspondingly we introduce three ``momenta": $\kappa_{\rm L}$, $\kappa_{\rm S}$ and $\kappa_{\rm sc}$, which give a FS involving $(n_{\rm c}+1)/2$, $n_{\rm c}/2$, and $n_{\rm c}$ electrons, respectively.\\
{\it Region I} ($0\leq H<H_1$):
Without magnetic field, $\mu_{\sigma}-{\rm Re}\Sigma_{\sigma}(0)$ corresponds to $\kappa_{\rm L}$ and the FS involves localized spins. 
Upon turning on the magnetic field, the FS splits, but the average keeps at $\kappa_{\rm L}$. \\
{\it Region II} ($H_1<H<H_2$):
At $H_1 \simeq 0.006$, one band passes the 
band edge
$\kappa_{\rm c}=1$ indicating that the chemical potential is located in the energy gap.
Then, only the other spin forms the FS, which has the momentum $\kappa_{\rm sc}$. 
Namely, the FS of a single spin-component involves $n_{\rm c}$ electrons in this region,
and the magnetization shows the plateau.
\\
{\it Region III} ($H>H_2$):
By further increasing the magnetic field, both spin components again contribute to the FS.
For $H>H_2 \simeq 0.126$, the spin average of $\mu_{\sigma}-{\rm Re}\Sigma_{\sigma}(0)$ is located at $\kappa_{\rm S}$,
 indicating the small FS composed of both spin-components.

We note that even in Region III, the quasiparticle involves the localized spins near $H_2$, since the energy gap cannot be formed without coupling with localized spins in the Kondo lattice. 
Namely, the ``small FS" does {\em not} mean the localized character of 4f electrons\cite{Miyake-Ikeda}. 
The energy gap gradually vanishes with decreasing 4f-contribution to the quasiparticle for $H \gg H_2 \sim T_{\rm K}$.

Transition of the FS topology is known as Lifshitz transition, which is distinguished from thermodynamic phase transition\cite{Lifshitz, Blanter94}.
The transitions between Regions I and II, or II and III in the Kondo lattice 
are identified as 
the Lifshitz transition in the sense that 
there is no discontinuity in physical quantities at finite temperature. 
However, as important characteristics in the Kondo lattice, the transition is caused by quasiparticles, which incorporate strong local correlation.
This difference appears in the renormalization factor $z_{\sigma}$ defined by
\begin{align}
	z_{\sigma} 
	= \left( 1 - \left. \frac{\partial {\rm Im}\Sigma_{\sigma}({\rm i}\epsilon_n)}{\partial \epsilon_n}
	\right|_{\epsilon_n \to +0}  \right)^{-1},
\end{align}
inverse of which corresponds to the mass enhancement of the quasiparticle.
Since now the energy gap is due to the collective Kondo singlet, ${\rm Im}\Sigma(\omega)$ has a peak centered at the energy gap as in the Mott insulator\cite{Georges}. 
This peak leads to a steep slope in the real part ${\rm Re}\Sigma(\omega)$.
As a result, the renormalization factor takes a small value when the quasiparticle level crosses the Fermi level, as shown in Fig.~\ref{fig:H_depend}(c).
Consequently, the effective mass for the majority spin is enhanced toward the magnetic field where the FS transition occurs,
while the other spin, which is not involved in the FS transition, 
becomes lighter
on the plateau region.
The effective mass strongly depends on the spin component on both sides of the transition, 
while the spin dependence is negligible for $H \gg H_2$.
This enhancement has been discussed 
in the high-field region\cite{Wasserman, Edwards-Green, Otsuki-JMMM}.

\section{Effect of Disorder}
\label{sec:disorder}

\begin{figure}[tb]
	\begin{center}
	\includegraphics[width=7.5cm]{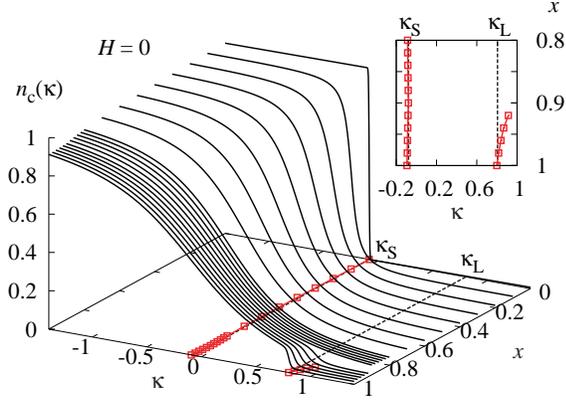}
	\end{center}
	\caption{(Color online) 
Momentum distribution for $H=0$.
The temperature is taken as
$T=0.001$ for $0.9 \leq x \leq 1$ and $T=0.0025$ for $0 \leq x \leq 0.88$,
which are low enough to neglect the broadening by finite temperature in the scale of the figure.
The (red) squares on the bottom and in the inset show the momenta where $-{\rm d}n_{\rm c}(\kappa)/{\rm d}\kappa$ 
takes a local maximum.}
	\label{fig:nk}
\end{figure}

Now let us examine the effect of disorder on the FS. 
We discuss the momentum distribution of conduction electrons
\begin{align}
	\langle c_{\bm{k}\sigma}^{\dag} c_{\bm{k}\sigma} \rangle
	= T\sum_n G_{\sigma}(\kappa, {\rm i}\epsilon_n) {\rm e}^{{\rm i}\epsilon_n 0+}
	\equiv n_{{\rm c}\sigma} (\kappa).
\end{align}
Figure~\ref{fig:nk} shows $n_{{\rm c}+}(\kappa) = n_{{\rm c}-}(\kappa) \equiv n_{\rm c}(\kappa)$ for $H=0$. 
Concerning $0 \leq x \leq 0.9$, 
the temperature is much lower than $T_{\rm K }$, and the results
may be regarded as those in the ground state.
In the case of $x=0$, $n_{\rm c}(\kappa)$ has a discontinuity of 
unit magnitude 
according to the Fermi-Dirac distribution function.
The Fermi momentum is determined only by the number of conduction electrons.
At $x=1$, on the other hand, a small discontinuity appears at the momentum where the Fermi volume includes localized spins\cite{Otsuki-LFS}. 
This large FS disappears at $x=0.9=n_{\rm c}$.
The volume of the large FS 
increases by removing localized spins toward $x=n_{\rm c}$.
This trend is in contrast to a rigid band picture that the volume decreases in proportion to the number of localized spins. 
It can be understood by considering that $x=n_{\rm c}$ is a condition for insulator. 
Namely, the Fermi momentum varies toward $\kappa_{\rm c}=1$ at $x=n_{\rm c}$.

\begin{figure}[tb]
	\begin{center}
	\includegraphics[width=7.5cm]{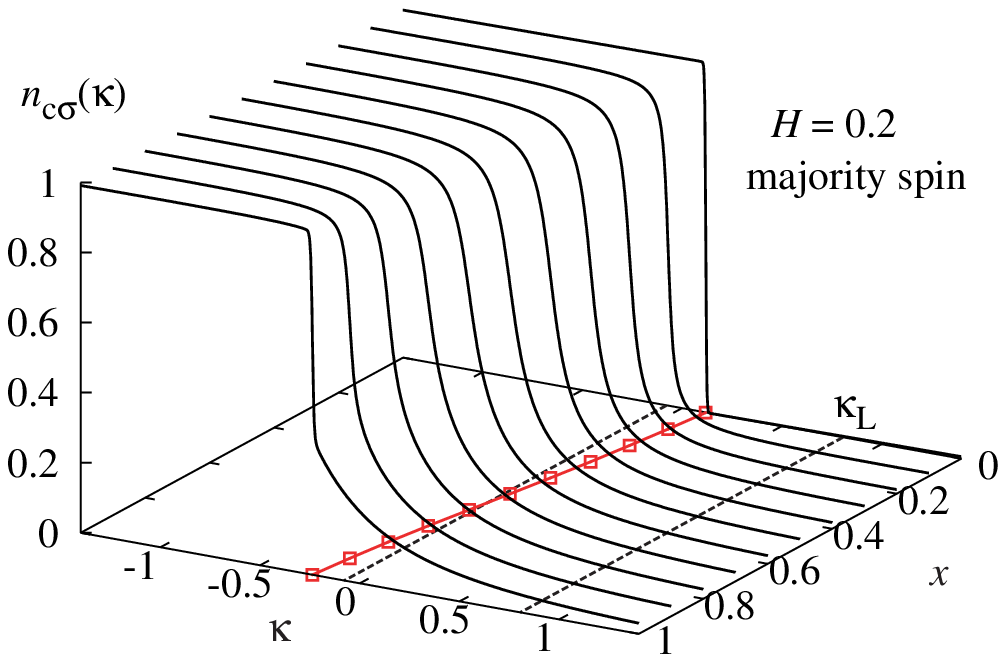}
	\includegraphics[width=7.5cm]{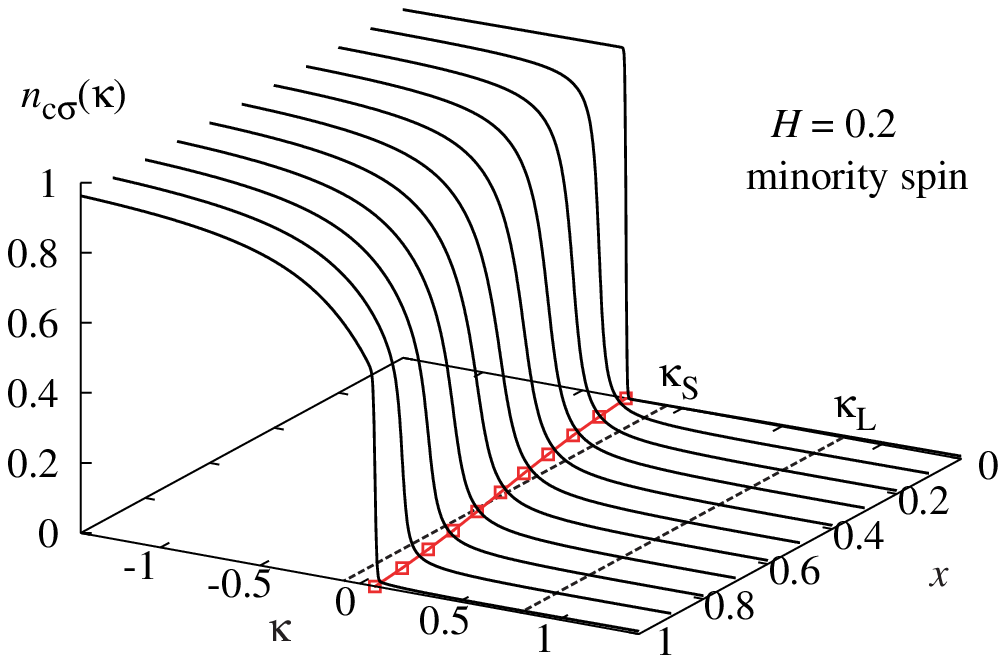}
	\end{center}
	\caption{(Color online) Momentum distribution $n_{{\rm c}\sigma}(\kappa)$ for $H=0.2$ and $T=0.0025$. Upper and lower panels correspond to majority and minority spins, respectively. The (red) squares on the bottom show the momentum where $-{\rm d}n_{{\rm c}\sigma}(\kappa)/{\rm d}\kappa$ 
takes the maximum.}
	\label{fig:nk-H}
\end{figure}

For magnetic field of $H_1< H \lesssim H_2$, 
$n_{{\rm c}\sigma}(\kappa)$ depends on $x$ in a similar manner as that in Fig.~\ref{fig:nk}, 
except that the FS in the Ce side appears only in the minority spin.
On the other hand,
when $H \gg H_2$, 
the FS can be connected continuously from $x=0$ to $x=1$.
Figure~\ref{fig:nk-H} shows $n_{{\rm c}\sigma}(\kappa)$ for $H=0.2$, which is much larger than $H_2$ and 
hence much larger than $T_{\rm K}$.
The FS's of majority and minority spins split as the Ce concentration increases.
The size of each component is opposite to the corresponding number of local spins 
 because of the effective field through the exchange interaction.
 
\begin{figure}[tb]
	\begin{center}
	\includegraphics[width=8cm]{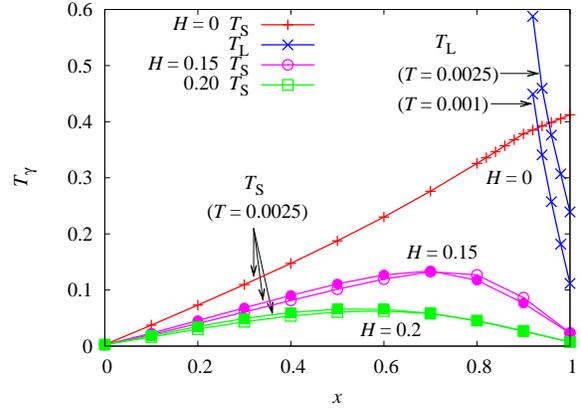}
	\end{center}
	\caption{(Color online) 
Dingle temperature $T_{\gamma}$ as a function of $x$ for several values of $H$. For finite $H$, the open and closed symbols represent majority and minority spins, respectively.}
	\label{fig:dingle}
\end{figure}

We define an energy scale $T_{\gamma \sigma}$ that characterizes the scattering rate of the quasiparticles\cite{Otsuki-LFS}:
\begin{align}
	T_{\gamma \sigma} = \frac{1}{4} 
	\left( -\frac{{\rm d}n_{{\rm c}\sigma}(\kappa)}{{\rm d}\kappa} \right)^{-1}_{{\rm FS}(\gamma)},
\end{align}
where $\gamma={\rm S}, {\rm L}$, and the subscript FS($\gamma$) indicates that the derivative is evaluated at the corresponding FS.
At $x=0$, the FS gets blurred according to $T_{{\rm S} \sigma}=T$, since $n_{\rm c}(\kappa)$ is given by the Fermi-Dirac distribution function.
On the other hand, the FS at $x=1$ depends more strongly on temperature: $T_{{\rm L} \sigma}=T/z_{\sigma}^2$\cite{Otsuki-LFS}.
Except for $x=0$ and $x=1$, $T_{\gamma \sigma}$ is finite even at $T=0$ because of disorder. 
In this case, $T_{\gamma \sigma}$ can be regarded as the Dingle temperature.

Figure~\ref{fig:dingle} shows $T_{\gamma \sigma}$ as a function of $x$ for several values of $H$.
At $H=0$, $T_{\rm S}$ monotonously increases as $x$ increases.
Namely, the La FS is not connected to the Ce side, although it leaves a trace of the ``small FS".
Instead, the ``large FS" appears at $x \geq n_{\rm c}$, appearance of which is indicated by rapid decrease of $T_{\rm L}$.
On the other hand, when $H>H_2$, $T_{\rm S}$ becomes zero at both $x=0$ and $x=1$ having a maximum at $0.5 \lesssim x \lesssim 0.8$.
Namely, the FS in the La side 
remains a true FS up to the Ce end.

\section{Summary and Discussions}
\label{sec:discussion}

\begin{figure}[tb]
	\begin{center}
	\includegraphics[width=8cm]{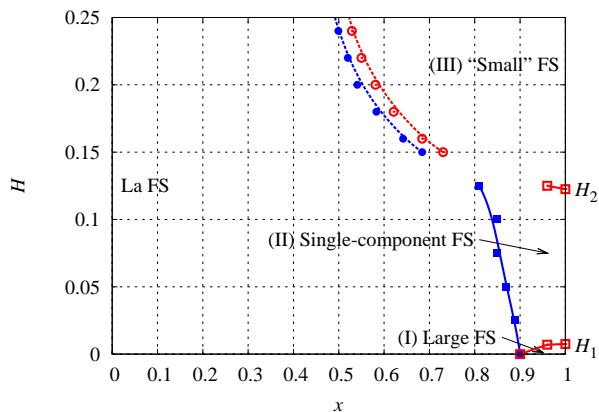}
	\end{center}
	\caption{(Color online) The FS topology in $x$-$H$ plane for $J=0.34$ and $n_{\rm c}=0.9$. The open square (red line) represents $H$ of metamagnetic behavior. The closed square (blue line) represents $x$ above which the indication of the quasiparticle is seen in $n_{{\rm c}\sigma}(\kappa)$. The dashed line and circle symbol represent $x$ where $T_{{\rm S}\sigma}$ 
takes the maximum. The open (red) and closed (blue) symbols signify anomaly due to majority and minority spins, respectively.}
	\label{fig:diagram}
\end{figure}

We summarize the above results in Fig.~\ref{fig:diagram}.
The Ce side is classified into three regions according to the FS topology: (I) large FS, (II) single-component FS, and (III) ``small'' FS.
This transition is caused by the quasiparticle band, whose energy gap $H_2-H_1$ is of the order of $T_{\rm K}$.
The effect of disorder on the FS 
is qualitatively different depending on 
whether $H$ is larger or smaller than $H_2$. 
When $H\lesssim H_2$, the heavy-fermion FS of one spin component appears around $x=n_{\rm c}$ upon increasing $x$.
In addition, if $H<H_1$, the other spin component shows up for larger $x$. 
In the special case of $H=0$, both spin components appear simultaneously at $x=n_{\rm c}$, at which the system becomes insulating. 
In this magnetic-field region, $H\lesssim H_2$, the Dingle temperature of the La FS monotonously increases against $x$.
On the other hand, 
when $H \gg H_2$,
the La FS is connected 
continuously 
to the Ce side.
In this case, the Dingle temperature has a maximum at a certain value of $x$.
As shown in Fig~\ref{fig:diagram}, the maximum point tends to converge at $x=0.5$ in high field, where the system can be regarded as a nonmagnetic binary alloy.

In realistic situations, the chemical potential is not in the hybridization gap, because of an energy dispersion of 4f level and/or existence of plural conduction bands on the Fermi level. 
In this case, a situation may occur where 
the magnetization-plateau region disappears in Fig.~\ref{fig:diagram}.
Then, we expect a metamagnetic behavior without the plateau around $H_1 \simeq H_2 \sim T_{\rm K}$.

We finally discuss realistic materials such as Ce$_x$La$_{1-x}$Ru$_2$Si$_2$.
We consider that the FS transition in pure CeRu$_2$Si$_2$ is the Lifshitz transition caused by the quasiparticle band\cite{Miyake-Ikeda}.
Actually, an analysis of transport coefficients indicates disappearance of a FS with one of spin components under magnetic field\cite{Daou06}.
Our numerical calculation demonstrates that this FS transition gives rise to a metamagnetic transition as shown in Fig.~\ref{fig:H_depend}(a).
In considering substitution effect in the multi-band system, the Dingle temperature is more informative rather than the dHvA frequencies, since each frequency may vary with keeping the total volume of the FS. 
Recent experiments have revealed a contrastive behavior depending on 
direction of
the magnetic field:
dHvA signals vanish by the substitution on both La and Ce sides for $\bm{H} \perp {\bf c}$,
while the dHvA amplitude $A_{\rm osc}$ shows a minimum at $0<x<1$ for $\bm{H} \parallel {\bf c}$\cite{Matsumoto-Aoki}.
Because the magnetization in this system shows a strong Ising anisotropy, we can regard the results for $\bm{H} \perp {\bf c}$ as low-field region, and the results for $\bm{H} \parallel {\bf c}$ as high-field region.
Then, the tendency is consistent with our results for the Dingle temperature $T_{\gamma}$, considering $A_{\rm osc} \propto \exp(-T_{\gamma})$.
The degree of $x$-dependence in the Dingle temperature indicates how much 4f electrons contribute to each branch.

%

\section*{Acknowledgments}
We thank H. Aoki and Y. Matsumoto for useful discussions.
This work was supported by a Grant-in-Aid for Scientific Research on Innovative Areas ``Heavy Electrons" (No. 20102008) of the Ministry of Education, Culture, Sports, Science, and Technology, Japan.

\end{document}